\documentclass[aps,prl,floatfix,twocolumn]{revtex4}
\usepackage{setspace}
\usepackage{graphics}
\usepackage{dcolumn}
\usepackage{epsfig}
\usepackage{array}
\usepackage{multirow}
\usepackage{amsmath}

\newcommand{\EP}{{\it e}-ph}
\newcommand{\EE}{{\it e}-{\it e}}
\newcommand{\RHO}{\rho_{e{\mbox-}{\rm ph}}}

\begin{document}

\bibliographystyle{prsty}

\title
{Electron-phonon interactions and the intrinsic electrical resistivity of graphene}
\author{Cheol-Hwan Park$^{1,2,3\ddagger}$}
\email{cheolhwan@snu.ac.kr}
\author{Nicola Bonini$^{4\ddagger}$}
\email{nicola.bonini@kcl.ac.uk}
\author{Thibault Sohier$^5$}
\author{Georgy Samsonidze$^{2}$}
\author{Boris Kozinsky$^{2}$}
\author{Matteo Calandra$^5$}
\author{Francesco Mauri$^5$}
\author{Nicola Marzari$^{1}$}
\affiliation{$^1$Theory and Simulation of Materials, \'Ecole Polytechnique
F\'ed\'erale de Lausanne, 1015 Lausanne, Switzerland\\
$^2$Research and Technology Center, Robert Bosch LLC, Cambridge, MA 02142, USA\\
$^3$Department of Physics and Astronomy
and Center for Theoretical Physics, Seoul National University, Seoul 151-747, Korea\\
$^4$Department of Physics, King's College London, London WC2R 2LS, UK\\
$^5$IMPMC, Universit\'e Pierre et Marie Curie, CNRS, 4 Place Jussieu, 75005 Paris, France\\
($^\ddagger$These authors contributed equally to this work.)}

\date{\today}

\begin{abstract}
We present a first-principles study of the temperature- and density-dependent intrinsic 
electrical resistivity of graphene.
We use density-functional theory and density-functional perturbation
theory together with very accurate Wannier interpolations
to compute all electronic and vibrational properties and electron-phonon
coupling matrix elements; the phonon-limited resistivity is then calculated within
a Boltzmann-transport approach.
An effective tight-binding model, validated against first-principles results, is also
used to study the role of electron-electron interactions
at the level of many-body perturbation theory.
The results found are in excellent agreement with
recent experimental data on graphene samples
at high carrier densities and elucidate the role of the different phonon modes
in limiting electron mobility.
Moreover, we find that the resistivity arising from scattering
with transverse acoustic phonons is 2.5 times higher than
that from longitudinal acoustic phonons.
Last, high-energy, optical and zone-boundary phonons contribute
as much as acoustic phonons to the intrinsic electrical resistivity
even at room temperature and become dominant at higher temperatures.
\end{abstract}
\maketitle

The intrinsic electrical resistivity $\RHO$ of graphene
arising from electron-phonon (\EP) interactions
provides a textbook example of carrier dynamics
in two dimensions~\cite{Physics3.106}: indeed,
$\RHO$ is proportional to $T^4$ at low temperatures, while at high temperatures $\RHO$
varies linearly with $T$
and, quite remarkably, is independent of doping.
The transition between these two distinct regimes is determined by the Bloch-Gr\"uneisen
temperature $\Theta_{\rm BG}=2\hbar\,v_s\,k_{\rm F}/k_{\rm B}$, where $\hbar$ and $k_{\rm B}$ are
the Planck and Boltzmann constants, $v_s$ the sound velocity, $k_{\rm F}$ the Fermi wavevector
(in case of graphene, measured at one of the two Dirac points ${\bf K}$ and ${\bf K}'$).
This characteristic temperature  $\Theta_{\rm BG}$,
as a result of its dependence on $k_{\rm F}$, is highly
tunable by changing gate voltages.
This scenario, first introduced theoretically by Hwang and Das\,Sarma~\cite{PhysRevB.77.115449},
has been confirmed experimentally by Efetov and Kim
using graphene samples at ultrahigh carrier densities~\cite{PhysRevLett.105.256805}.

In spite of this clear picture, there are, however, several open questions.
For instance, contrary to the expected high-temperature behavior~\cite{PhysRevB.22.904,PhysRevB.77.115449},
a significant {\it charge-density-dependent} nonlinear behavior in
$\RHO(T)$ has been reported~\cite{ywtan:euro,PhysRevLett.100.016602,chen:nnano,PhysRevLett.101.096802,
PhysRevLett.104.236601,PhysRevLett.105.266601};
nonlinearities are found to be stronger when the charge density
is lower~\cite{chen:nnano,PhysRevLett.101.096802}.
The origin of this behavior is not clearly understood yet, and explanations involve
temperature-dependent screening in graphene~\cite{PhysRevB.79.165404},
substrate surface phonons~\cite{PhysRevB.77.195415,chen:nnano},
or rippling and flexural phonons~\cite{PhysRevLett.100.016602,PhysRevLett.100.076801,PhysRevB.82.195403,PhysRevLett.105.266601}.

For the resistivity at high densities, currently there is no formulation
that explains the experimental $\RHO$ reported in Refs.~\cite{chen:nnano,PhysRevLett.105.256805}
without any fitting parameters.
Moreover, the relative role of the different acoustic and optical
phonon modes has not been elucidated yet, as well as the detailed nature of the electron-acoustic phonon interactions.
In particular, in 1980 Pietronero {\it et al.}~\cite{PhysRevB.22.904} derived the
high-temperature ($T\gg\Theta_{\rm BG}$) limit for $\RHO(T)$ considering the contribution
of the {\it gauge field} (arising from the changes in the local electronic hopping integrals, due to bond-length variations)
to $\RHO$ of graphene for both longitudinal acoustic (LA) and transverse acoustic
(TA) phonon modes.
More recently, it has been argued~\cite{PhysRevB.61.10651,PhysRevB.65.235412} that,
in addition to this gauge-field contribution,
a {\it deformation-potential} contribution should be considered to properly account for $\RHO$.
This contribution would only be relevant for LA phonons~\cite{PhysRevB.76.045430} and proportional to the local
electron-density change upon deformation (we note in passing that
in some cases, e.\,g.\,, in Refs.~\cite{PhysRevB.61.10651,PhysRevB.81.195442},
the term `deformation potential' has been used to denote
what is labeled `gauge field' in this paper and in other works, e.\,g.\,, Ref.~\cite{PhysRevB.65.235412}).

The relative importance of these gauge-field and deformation-potential contributions
to $\RHO$ is currently heavily debated.
Woods and Mahan~\cite{PhysRevB.61.10651} estimate that the gauge-field term
is $\sim20$ times [$=(3.87/0.87)^2$] more important than the deformation-potential term
in determining $\RHO$; Von Oppen, Guinea and Mariani~\cite{PhysRevB.80.075420} argue
that the deformation potential contribution to the \EP\ coupling matrix elements
is negligible in comparison with the gauge-field term for small-wavevector scattering, and
various authors have used this assumption in the calculation of $\RHO$~\cite{PhysRevB.76.205423,PhysRevB.81.195442,PhysRevB.82.195403}.
On the other hand, Suzuura and Ando~\cite{PhysRevB.65.235412} suggest that
the contribution of the deformation potential to $\RHO$ is much more important than
the gauge-field contribution and estimate the ratio of the two
to be $\sim(30/1.5)^2=400$.
Based on the assumption that the gauge field and TA phonon modes
are not important, Hwang and Das\,Sarma~\cite{PhysRevB.77.115449}
have modeled $\RHO$ considering only LA phonons and with
an {\it effective} deformation potential
where all the complex
dependence of the \EP\ coupling matrix elements on electron and phonon
wavevectors is condensed into a single fitting parameter.

Recently, there have been attempts to calculate $\RHO$ based on models
of the \EP\ coupling matrix elements
fitted to first-principles calculations~\cite{PhysRevB.81.121412,PhysRevB.85.165440}.
The resistivities found, arising from acoustic phonons in the linear regime
($\RHO\propto T$) and reported in Ref.~\cite{PhysRevB.81.121412}
and in Ref.~\cite{PhysRevB.85.165440}, are~$\sim4$
and $\sim13$~times lower, respectively, than the experimental
results~\cite{chen:nnano,PhysRevLett.105.256805}.
High-energy optical phonons were
considered in Ref.~\cite{PhysRevB.81.121412} and their importance in the
high-temperature regime was underlined.

The main purpose of this paper is to provide a fully microscopic and first-principles characterization of
the temperature- and density-dependent phonon-limited electrical resistivity $\RHO$ in graphene.
We first use density-functional theory (DFT) and density-functional perturbation
theory (DFPT) as implemented in the {\tt Quantum-ESPRESSO} distribution~\cite{gianozzi:qe}
within the local-density approximation (LDA)~\cite{PhysRevLett.45.566,PhysRevB.23.5048}
to compute the electronic and vibrational properties including the \EP\ coupling matrix elements.
Next, we use these results to calculate the resistivity within a Boltzmann transport framework.
Then, first-principles
results are also used to validate an effective and accurate model for \EP\ couplings that includes
gauge-field and deformation-potential contributions.
This model allows the treatment of the effects of electron-electron (\EE) interactions
at the level of many-body perturbation theory on $\RHO$, which are discussed in detail.

The key ingredients to compute $\RHO$ are the \EP\ coupling matrix elements
\begin{equation}
g^\nu_{m',\,m}({\bf p},{\bf q})=\left<m',{\bf p}+{\bf q}\right|\Delta V^\nu_{{\bf q}}\left|m,{\bf p}\right>\,,
\label{eq:g}
\end{equation}
where $\left|m,{\bf p}\right>$ is an electronic eigenstate computed within DFT for a Bloch state with
energy $\varepsilon_{m,{\bf p}}$ (band index $m$ and wavevector ${\bf p}$), and $\Delta V^\nu_{\bf q}$
is the change in the self-consistent potential induced by a phonon mode with
energy $\hbar\omega^\nu_{\bf q}$ (branch index $\nu$ and wavevector {\bf q}).

Employing a first-principles interpolation scheme~\cite{PhysRevB.76.165108}
based on maximally localized Wannier functions~\cite{PhysRevB.56.12847,PhysRevB.65.035109,marzari:rmp},
as implemented in the {\tt Wannier90}~\cite{mostofi:cpc} and {\tt EPW}~\cite{jesse:CPC} packages,
we are able to calculate the electronic energies $\varepsilon_{m,{\bf p}}$, the band velocities
${\bf v}_{m,{\bf p}}\equiv\nabla_{\bf p}\,\varepsilon_{m,{\bf p}}/\hbar$,
the phonon frequencies $\omega^\nu_{\bf q}$ and the \EP\ coupling matrix elements
$g^\nu_{m',\,m}({\bf p},{\bf q})$ for {\bf p} and {\bf q} on ultra-dense grids
spanning the entire Brillouin zone,
crucial for an accurate and efficient evaluation of
the transport Eliashberg function~\cite{PhysRevB.17.3725}
\begin{eqnarray}
&&\alpha^2_{\rm tr}F(\omega)=\frac{1}{N_\uparrow}\sum_{m'\,m\,\nu}
\int\int \frac{d{\bf p}}{A_{\rm BZ}}\,\frac{d{\bf q}}{A_{\rm BZ}}\,
|g^\nu_{m',\,m}({\bf p},{\bf q})|^2
\nonumber\\
&&\times\,
\delta(\varepsilon_{m',{\bf p}+{\bf q}}-E_{\rm F})\,
\delta(\varepsilon_{m,{\bf p}}-E_{\rm F})\,
\delta(\hbar\omega^\nu_{\bf q}-\hbar\omega)
\nonumber\\
&&\times\,
\left(1-\frac{{\bf v}_{{\bf p}+{\bf q},m'}\,\cdot\,{\bf v}_{{\bf p},m} }{|{\bf v}_{{\bf p},m}|^2}\right)
\label{eq:a2tr}
\end{eqnarray}
($N_\uparrow$ is the density of states per
spin per unit cell at $E_{\rm F}$). Each integration extends over
the Brillouin zone of graphene, of area $A_{\rm BZ}=8\sqrt{3}\,\pi^2/9\,b^2$,
where $b$ is the carbon-carbon bond length.
We finally evaluate $\RHO$
using the lowest-order variational solution of the Boltzmann transport equation~\cite{PhysRevB.17.3725}:
\begin{equation}
\RHO=\frac{3\sqrt{3}\,\pi\,b^2}{e^2\,N_\uparrow\,v^2}\int_0^\infty
\frac{\hbar\omega/2k_{\rm B}T}{\sinh^2(\hbar\omega/2k_{\rm B}T)}\,
\alpha_{\rm tr}^2F(\omega)\,d\omega\,,
\label{eq:RHO}
\end{equation}
where $e$ is the charge of an electron and $v^2$ is the electronic band velocity
squared and averaged over the Fermi surface.
Equations~(\ref{eq:a2tr}) and~(\ref{eq:RHO}) are based on the assumption that
the electronic density of states does not change appreciably near the Fermi level
over the phonon energy scale, which is always valid {\it either} (i) if the temperature
is lower than room temperature (so that acoustic phonons dominantly participate in electron
scattering) {\it or} (ii) if graphene is heavily doped.
[As an example, if
$E_{\rm F}>0.5$~eV (as measured from the Dirac point energy),
the product of the initial and scattered electron densities of states
is proportional to $(E_{\rm F}+\hbar\omega/2)(E_{\rm F}-\hbar\omega/2)$,
instead of $E_{\rm F}^2$, resulting in an error $<4\%$ in $\RHO$
arising from optical phonons with $\hbar\omega=0.2$~eV.]

Technical details of the calculations are as follows.
A kinetic energy cutoff of 60~Ry is used in expanding the valence electronic
states in a planewave basis~\cite{ihm:1979JPC_PW}, and core-valence interactions
are taken into account by means of norm-conserving
pseudopotentials~\cite{PhysRevB.43.1993}.
Charge doping is modeled by adding extra electrons and a neutralizing background.
The bond length in the calculations is $b=1.405$~\AA\ (for intrinsic graphene)
and each graphene sheet is separated from its periodic replicas by 8.0~\AA\
to ensure that the effect of periodic boundary conditions are negligible.
We have used Brillouin zone integrations of $72\times72\times1$ {\bf p} points
in the full Brillouin zone
for all charge density and phonon calculations.
All quantities $\varepsilon_{m,{\bf p}}$, ${\bf v}_{m,{\bf p}}$,
$\omega^\nu_{\bf q}$, and $g^\nu_{m',\,m}({\bf p},{\bf q})$
have been calculated first for {\bf p} or {\bf q}
on a coarse grid of $6\times6\times1$ points in the full Brillouin zone
and then Wannier interpolated into a fine
grid of $300\times300\times1$ points {\it in the irreducible wedge}.
Lorentzians with a finite broadening of 0.025~eV
were used for the two energy delta
functions involving electronic energies $\varepsilon_{m,{\bf p}}$
and $\varepsilon_{m',{\bf p}+{\bf q}}$ in Eq.~(\ref{eq:a2tr});
such an approximation is not necessary for the delta function
involving $\omega^\nu_{\bf q}$
thanks to the integration over $\omega$ in Eq.~(\ref{eq:RHO}).

\begin{figure}
\begin{center}
\includegraphics[width=1.0\columnwidth]{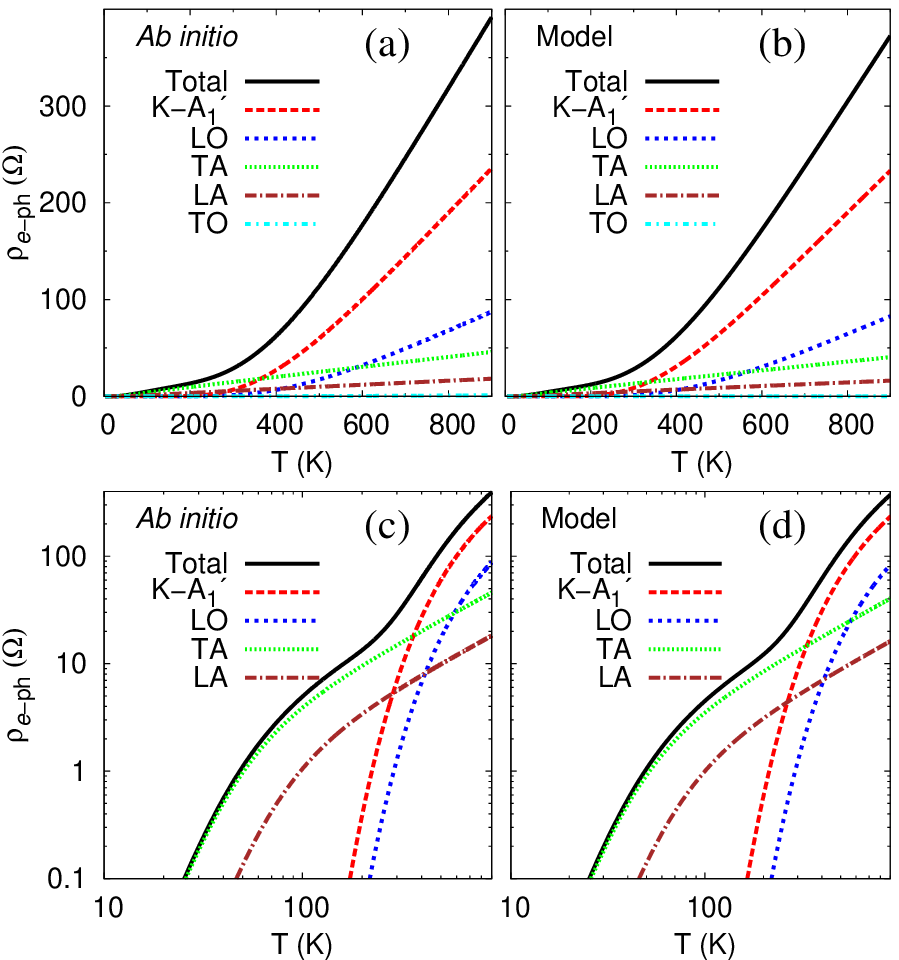}
\end{center}
\caption{(a) and (c): Electrical resistivity of {\it n}-doped
graphene ($n=2.86\times10^{13}$~cm$^{-2}$) arising from \EP\ interactions
versus the temperature calculated from first principles within the LDA.
The partial resistivity arising from each phonon branch is also shown.
(b) and (d): Similar quantities as in (a) and (c) obtained from
the model calculation based on the LDA results (see text).
Quantities are plotted in linear scale in (a) and (b) and in logarithmic
scale in (c) and (d).}
\label{Fig1}
\end{figure}

\begin{figure}
\begin{center}
\includegraphics[width=1.0\columnwidth]{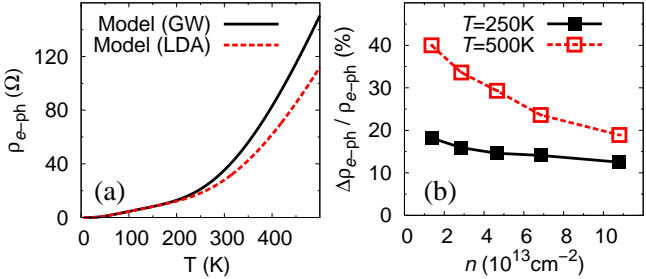}
\end{center}
\caption{(a) Electrical resistivity of {\it n}-doped
graphene ($n=2.86\times10^{13}$~cm$^{-2}$) arising from \EP\ interactions
as a function of temperature, as calculated from our model with
\EE\ interaction effects taken into account at the {\it GW} level~\cite{PhysRev.139.A796}
(solid or black curve)
or using LDA (dashed or red curve).
(b) Relative change in the calculated $\RHO$ versus $n$
due to \EE\ interaction effects.}
\label{FigGW}
\end{figure}

\begin{figure}
\begin{center}
\includegraphics[width=1.0\columnwidth]{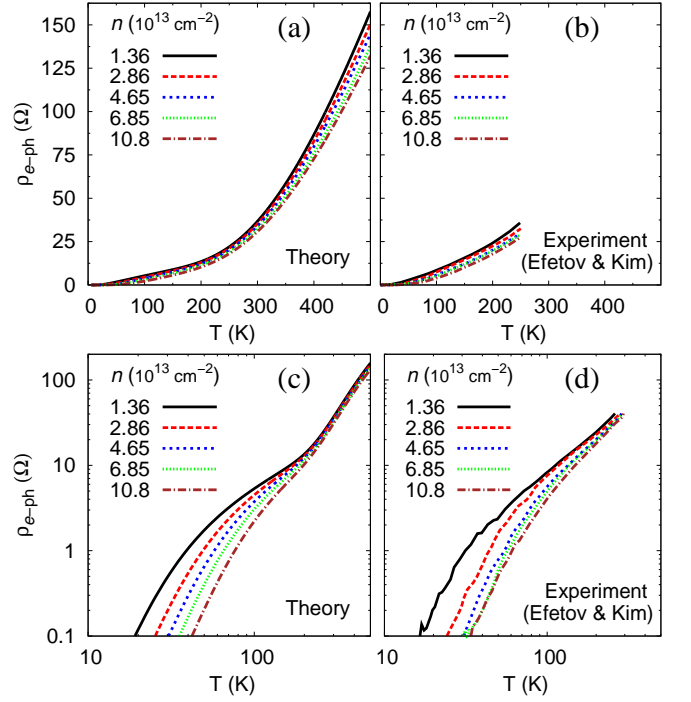}
\end{center}
\caption{Electrical resistivity of {\it n}-doped
graphene arising from \EP\ interactions
versus temperature calculated from
the model (Tab.~\ref{Tab3}) incorporating \EE\ interaction effects
[(a) and (c)] and the corresponding experimental data
from Ref.~\cite{PhysRevLett.105.256805} [(b) and (d)].
Quantities are plotted in linear scale in (a) and (b) and in logarithmic
scale in (c) and (d).}
\label{Fig3}
\end{figure}

\begin{table*}
\caption{Our suggested model for the {\it e}-ph coupling matrix element
$g^\nu_{m',\,m}({\bf p},{\bf q})=\left<m',{\bf p}+{\bf q}\right|\Delta V^\nu_{{\bf q}}\left|m,{\bf p}\right>$,
considering nearest-neighbor electron hoppings and ion-ion interactions
in graphene and assuming that $|{\bf k}|\ll|{\bf K}|$ and  $|{\bf k}'|\ll|{\bf K}|$.
The electronic band indices ($m$ and $m'$) are $+1$ and $-1$ for the upper and lower bands, respectively.
The angle $\theta_{\bf c}$ for a two-dimensional vector~{\bf c} is the polar angle between
{\bf c} and the $\Gamma$K direction.
The angles $\theta_{\bf k}$ and $\theta_{{\bf k}'}$ are
set to be in the interval $[0,2\pi)$, without losing generality.
Here, $b$ ($=1.405$~\AA) is the relaxed carbon-carbon bond length,
$v_s^{\rm TA}$ ($=14.2$~km/s) and $v_s^{\rm LA}$ ($=22.4$~km/s) the first-principles sound velocities
for the TA and LA phonon modes, respectively, and $M_{\rm C}$ the mass of a carbon atom.
The calculated phonon energies $\hbar\omega^{{\rm E}_{2g}}_\Gamma$ ($=0.20$~eV) and $\hbar\omega^{{\rm A}'_1}_{\bf K}$
are for degenerate E$_{2g}$ modes ($\nu=5,\,6$) at
${\bf q}=\Gamma$ and for the A$_1'$ mode ($\nu=6$) at
${\bf q}={\bf K}$, respectively.
If modeling LDA results (Fig.~\ref{Fig1}
and Figs.~\ref{SFig2} and~\ref{SFig1}),
we set $v=v_{\rm LDA}=0.866\times10^6$~m/s,
the hopping integral $|\gamma|=|\gamma_{\rm LDA}|=\frac{2\hbar}{3b}\,v_{\rm LDA}=2.67$~eV,
the coupling parameter $\eta=\eta_{\rm LDA}=\frac{|\gamma_{\rm LDA}|}{b}
\left(1-\frac{b}{v_{\rm LDA}}\frac{d v_{\rm LDA}}{db}\right)=4.75$~eV\,\AA$^{-1}$,
and $\hbar\omega^{{\rm A}'_1}_{\bf K}=0.160$~eV.
See text for the modeling of \EE\ interaction effects beyond the LDA.
Finally, the parameter $D$ defines the strength of the deformation potential,
which is set to zero here.}
\renewcommand{\arraystretch}{2.0}
\begin{tabular}{ c | c}
\hline
\hline
{\it E}-ph coupling matrix element &  Model\\
\hline
\hline
$\left<+1,{\bf K}+{\bf k}'\right|\Delta V_{{\bf k}'-{\bf k}}^{\rm TA}\left|+1,{\bf K}+{\bf k}\right>$&
$
\sqrt{\frac{\hbar\,|{\bf k}'-{\bf k}|}{4\,M_{\rm C}\,v_s^{\rm TA}}}\,
\frac{3}{4}\,
b\,\eta\,
\sin\left(\frac{\theta_{{\bf k}}+\theta_{{\bf k}'}}{2}+2\,\theta_{{\bf k}'-{\bf k}}\right)$ \\
If $|{\bf k}'|=|{\bf k}|$
&
$-\sqrt{\frac{\hbar\,|{\bf k}'-{\bf k}|}{4\,M_{\rm C}\,v_s^{\rm TA}}}\,
\frac{3}{4}\,
b\,\eta\,
\sin\frac{3}{2}(\theta_{{\bf k}}+\theta_{{\bf k}'})$\\
\hline
$\left<+1,{\bf K}+{\bf k}'\right|\Delta V_{{\bf k}'-{\bf k}}^{\rm LA}\left|+1,{\bf K}+{\bf k}\right>$&
$i\,
\sqrt{\frac{\hbar\,|{\bf k}'-{\bf k}|}{4\,M_{\rm C}\,v_s^{\rm LA}}}\,
\left[
\frac{3}{4}\,b\,\eta\,
\cos\left(\frac{\theta_{{\bf k}}+\theta_{{\bf k}'}}{2}+2\,\theta_{{\bf k}'-{\bf k}}\right)
+D\,\cos\frac{\theta_{\bf k}-\theta_{{\bf k}'}}{2}
\right]$ \\
If $|{\bf k}'|=|{\bf k}|$
&
$i\,
\sqrt{\frac{\hbar\,|{\bf k}'-{\bf k}|}{4\,M_{\rm C}\,v_s^{\rm LA}}}\,
\left[
-\frac{3}{4}\,b\,\eta\,
\cos\frac{3}{2}(\theta_{{\bf k}}+\theta_{{\bf k}'})
+D\,\cos\frac{\theta_{\bf k}-\theta_{{\bf k}'}}{2}
\right]$ \\
\hline
$\left<+1,{\bf K}+{\bf k}'\right|\Delta V_{{\bf k}'-{\bf k}}^{\rm TO}\left|+1,{\bf K}+{\bf k}\right>$&
$\sqrt{\frac{\hbar}{4\,M_{\rm C}\,\omega_{\Gamma}^{{\rm E}_{2g}}}}\,
3\,\eta\,
\cos\left(\frac{\theta_{{\bf k}}+\theta_{{\bf k}'}}{2}-\theta_{{\bf k}'-{\bf k}}\right)$ \\
If $|{\bf k}'|=|{\bf k}|$
&
0\\
\hline
$\left<+1,{\bf K}+{\bf k}'\right|\Delta V_{{\bf k}'-{\bf k}}^{\rm LO}\left|+1,{\bf K}+{\bf k}\right>$&
$-\sqrt{\frac{\hbar}{4\,M_{\rm C}\,\omega_{\Gamma}^{{\rm E}_{2g}}}}\,
3\,\eta\,
\sin\left(\frac{\theta_{{\bf k}}+\theta_{{\bf k}'}}{2}-\theta_{{\bf k}'-{\bf k}}\right)$ \\
If $|{\bf k}'|=|{\bf k}|$
&
$\begin{cases}
-\sqrt{\frac{\hbar}{4\,M_{\rm C}\,\omega_{\Gamma}^{{\rm E}_{2g}}}}\,
3\,\eta&
\mbox{if }\theta_{{\bf k}'}>\theta_{\bf k}\\
\sqrt{\frac{\hbar}{4\,M_{\rm C}\,\omega_{\Gamma}^{{\rm E}_{2g}}}}\,
3\,\eta&
\mbox{if }\theta_{{\bf k}'}<\theta_{\bf k}\\
\end{cases}$\\
\hline
$\left<+1,2{\bf K}+{\bf k}'\right|\Delta V_{{\bf K}+{\bf k}'-{\bf k}}^{{\rm A}_1'}\left|+1,{\bf K}+{\bf k}\right>$&
$i\,
\sqrt{\frac{\hbar}{4\,M_{\rm C}\,\omega_{\bf K}^{{\rm A}_1'}}}\,
3\sqrt{2}\,\eta\,
\sin\frac{\theta_{{\bf k}}-\theta_{{\bf k}'}}{2}$ \\
\hline
$\left<+1,2{\bf K}+{\bf k}'\right|\Delta V_{{\bf K}+{\bf k}'-{\bf k}}^{\nu=3,4,5}\left|+1,{\bf K}+{\bf k}\right>$&
0 (All three modes contribute $\le1$\,\% to the resistivity.)\\
\hline
Out-of-plane phonon modes & 0 ({\it E}-ph coupling matrix element is zero.)\\
\hline
If $m=-1$ & Replace ${\bf k}$ in the corresponding expression above by $-{\bf k}$.\\
If $m'=-1$ & Replace ${\bf k}'$ in the corresponding expression above by $-{\bf k}'$.\\
\hline
$\left<m',2{\bf K}+{\bf k}'\right|\Delta V_{{\bf k}'-{\bf k}}^{\nu}\left|m,2{\bf K}+{\bf k}\right>$&
$\left<m',{\bf K}-{\bf k}'\right|\Delta V_{{\bf k}-{\bf k}'}^{\nu}\left|m,{\bf K}-{\bf k}\right>^*$\\
$\left<m',{\bf K}+{\bf k}'\right|\Delta V_{-{\bf K}+{\bf k}'-{\bf k}}^{\nu}\left|m,2{\bf K}+{\bf k}\right>$&
$\left<m',2{\bf K}-{\bf k}'\right|\Delta V_{{\bf K}+{\bf k}-{\bf k}'}^{\nu}\left|m,{\bf K}-{\bf k}\right>^*$\\
\hline
\hline
\end{tabular}
\label{Tab3}
\end{table*}

\begin{figure*}
\begin{center}
\end{center}
\caption{
(Please contact the authors or visit Nano Letters web site for figures.)
Absolute value of the \EP\ coupling matrix elements
$\left|\left<+1,{\bf K}+{\bf k}'\right|\Delta V^\nu_{{\bf k}'-{\bf k}}\left|+1,{\bf K}+{\bf k}\right>\right|$
for LO and TO modes and
$\left|\left<+1,2{\bf K}+{\bf k}'\right|\Delta V^\nu_{{\bf K}+{\bf k}'-{\bf k}}\left|+1,{\bf K}+{\bf k}\right>\right|$
for in-plane phonon modes versus
${\bf k}'$ for graphene.
In each panel, the equi-energy states ($|{\bf k}'|=|{\bf k}|$) are denoted by a white circle.
For LO and TO modes, {\bf k} is represented by a green disk.
For ${\bf q}\sim{\bf K}$ in-plane phonon modes, {\bf k} is represented by a pink disk
and it should be noted that the Bloch wavevectors $2{\bf K}+{\bf k}'$ and ${\bf K}+{\bf k}$
are near the two (different) Dirac points $2{\bf K}$ and ${\bf K}$, respectively.
The first and second columns for each mode show first-principles
calculations within LDA on {\it n}-doped graphene
($n=2.86\times10^{13}$~cm$^{-2}$)
whose Fermi level lies at the equi-energy contour
and those on intrinsic graphene, respectively; the third column shows results
of the model calculations based on the LDA results (see Tab.~\ref{Tab3}).}
\label{SFig2}
\end{figure*}

The characteristic features of the phonon-limited resistivity in graphene
at high charge density are shown in Figs.~\ref{Fig1}(a) and~\ref{Fig1}(c).
Here we plot the total $\RHO(T)$ for {\it n}-doped graphene
(for a charge concentration of $2.86\times10^{13}$~cm$^{-2}$)
together with the contribution of the different phonon branches
(The contribution arising from the two phonon branches related
to out-of-plane vibrations is zero by symmetry, as pointed out in
Ref.~\cite{PhysRevB.76.045430}).
Within the LDA, for  $T<200$~K the resistivity is mainly due to
acoustic phonons, with the TA modes contributing $\sim$2.5~times more
than the LA ones.
Therefore, one can argue that a model for the \EP\ coupling matrix elements
which includes only deformation-potential
contributions that act on LA modes would not be fully adequate.
Interestingly, the slope increase at $T>200$~K is due to
the optical and zone-boundary phonons:
more specifically, it is mainly due to the A$_1'$ phonons near ${\bf q}={\bf K}$,
with a smaller contribution from longitudinal optical (LO)
phonons near ${\bf q}=\Gamma$.
Even at room temperature, 30\% of the
total $\RHO$ arises from these high-energy phonons.
When \EE\ interactions beyond the LDA are properly taken into account
(see Figs.~\ref{FigGW} and~\ref{Fig3} and the relevant discussion),
high-energy, optical and zone-boundary phonons are found to be as important
as acoustic phonons, accounting for 50\% of
the total $\RHO$ at room temperature,
and become dominant at higher temperatures.
[Equations~(\ref{eq:a2tr}) and (\ref{eq:RHO}) show that $\RHO$ is proportional
to the square of the \EP\ matrix element and is {\it inversely} proportional
to the square of the Fermi velocity.
If we take \EE\ interactions into account within the {\it GW} approximation,
the enhancement in the calculated \EP\ coupling matrix element
for the A$_1'$ branch near ${\bf q}={\bf K}$
is larger than that in the Fermi velocity~\cite{rubio:nl2010},
whereas the two enhancements almost cancel each other
for optical and acoustic phonons with small wavevectors
(we discuss this in more detail later).
In addition, the calculated energy of the zone-boundary A$_1'$ phonon
is reduced from its LDA value
by 7\% within the {\it GW} approximation~\cite{PhysRevB.78.081406,PhysRevB.80.085423}.
These two effects make the contribution to the calculated $\RHO$
of the high-energy phonons
in the {\it GW} approximation (50\% at room temperature)
larger than in LDA (30\% at room temperature).]
On the other hand, the contribution of transverse optical (TO)
phonons is negligible, since the \EP\ coupling matrix elements vanish
if the two electronic states involved in scattering have the
same energy (Tab.~\ref{Tab3} and Fig.~\ref{SFig2}).
We believe these results are very relevant for graphene electronic devices
operating at or above room temperature.

It is somewhat surprising that the optical and zone-boundary
phonons, whose energies are of the order of at least 0.150 eV,
which corresponds to 1740~K,
could contribute to $\RHO$ at room temperature (300~K)
as much as acoustic phonons.
This can be explained as follows.
First, in general, the crossover from the low-temperature
$\RHO$ vs.\,$T$ behavior (e.\,g.\,, $\RHO\propto T^5$ in
three dimensions) to the high-temperature one ($\RHO\propto T$) occurs
at a temperature which is 20\% and {\it not} 100\% of the relevant phonon
energy scale~\cite{matula}, although
the two energy scales (20\% and 100\% of the phonon energy)
are of the same order of magnitude.
In other words, high-energy phonons may contribute to
$\RHO$ at temperatures much lower than 1740~K
(e.\,g.\,, 20\% of 1740~K is 350~K).
Second, in the high-temperature regime ($\RHO\propto T$),
according to the model in Tab.~\ref{Tab3},
the ratio of the contribution to $\RHO$ from
high-energy phonons to that from acoustic phonons is
(see the caption of Tab.~\ref{Tab3} for the meaning of parameters used)
\begin{equation}
\frac{\RHO^{\rm high-energy}}{\RHO^{\rm acoustic}}=
\frac{
32\left[
\left(\omega^{{\rm E}_{2g}}_\Gamma\right)^{-2}
+\left(\omega^{{\rm A}'_1}_{\bf K}\right)^{-2}
\right]}
{b^2\left[
\left(v_s^{\rm LA}\right)^{-2}
+\left(v_s^{\rm TA}\right)^{-2}
\right]}\approx6.9\,,
\label{eq:highE_acoustic}
\end{equation}
which is much larger than 1; i.\,e.\,, high-energy
phonons are much more effective than acoustic phonons
in scattering electrons at high temperatures.
[In order to obtain Eq.~(\ref{eq:highE_acoustic}),
we used the model in Tab.~\ref{Tab3} and Eqs.~(\ref{eq:a2tr}) and (\ref{eq:RHO}).]

For a better understanding and application of our first-principles results,
we introduce a model based on nearest-neighbor electron hoppings and
lattice interactions that can provide \EP\ coupling matrix elements for all phonon branches
(acoustic, optical and zone-boundary) on an equal footing; this is of crucial importance to accurately
account for resistivity in a wide range of temperatures, as shown in Fig.~\ref{Fig1}(c).
A similar model (nearest-neighbor electron hoppings and lattice interactions) has been used in
Refs.~\cite{PhysRevLett.93.185503,ando:optical,PhysRevB.75.035427,ando:zoneboundary,PhysRevB.84.035433}
for the highest-energy E$_{2g}$ phonons at ${\bf q}=\Gamma$ and A$_1'$ phonons at ${\bf q}={\bf K}$,
while for acoustic phonons some studies have employed similar nearest-neighbor
interactions~\cite{PhysRevB.22.904,PhysRevB.84.035433}
(or variations including the
restoration torque for bending distortions~\cite{PhysRevB.61.10651,PhysRevB.65.235412})
to describe the gauge-field contribution to $\RHO$.

The parameters that enter the model are the energies of the E$_{2g}$ phonon at
${\bf q}=\Gamma$, $\hbar\omega_\Gamma^{{\rm E}_{2g}}$,
and of the A$_1'$ phonon at ${\bf q}={\bf K}$, $\hbar\omega_{\bf K}^{{\rm A}_1'}$,
the sound velocities of the TA ($v_s^{\rm TA}$) and LA ($v_s^{\rm LA}$) modes
and the coupling strength
$\eta=-\frac{d|\gamma|}{db}$, where
$|\gamma|=\frac{2\hbar}{3b}\,v$ is the absolute value of
the nearest-neighbor hopping integral (regarding the sign of $\gamma$,
which is relevant, e.\,g.\,, for photoemission experiments,
see Ref.~\cite{PhysRevB.84.125422}).
All these parameters are computed from first principles.
In particular, the \EP\ coupling term $\eta$
is obtained from the electronic band structure of isotropically
strained graphene,
$\eta=\frac{|\gamma|}{b}
\left(1-\frac{b}{v}\frac{d v}{db}\right)$.
The coupling strength is directly reflected in
the band velocity versus bond length relation;
an intermediate result necessary for calculating $\eta$
within the LDA, $\frac{b}{v}\frac{d v}{db}=-1.50$,
is found to be in good agreement with Ref.~\cite{PhysRevB.81.081407}.
$\eta$ has previously been obtained from comparison between
an analytical expression and first-principles results on the \EP\ coupling
strength (e.\,g.\,, Ref.~\cite{PhysRevLett.93.185503,ando:optical}).

\begin{figure*}
\begin{center}
\end{center}
\caption{
(Please contact the authors or visit Nano Letters web site for figures.)
Absolute value of the {\it renormalized} \EP\ coupling matrix element
$\left|\left<+1,{\bf K}+{\bf k}'\right|\Delta V^\nu_{{\bf k}'-{\bf k}}\left|+1,{\bf K}+{\bf k}\right>/
\sqrt{\frac{\hbar\,|{\bf k}'-{\bf k}|}{4\,M_{\rm C}\,v^\nu_s}}\right|$ versus
${\bf k}'$ for acoustic phonon branches ($\nu$ is either TA or LA) of graphene.
In each panel, {\bf k} is represented by a green disk and the equi-energy
states ($|{\bf k}'|=|{\bf k}|$) are denoted by a white circle.
The first and second columns for each mode show first-principles
calculations within LDA on {\it n}-doped graphene ($n=2.86\times10^{13}$~cm$^{-2}$)
whose Fermi level lies at the equi-energy contour and those on intrinsic graphene,
respectively; the third column shows results
of the model calculations based on the LDA results (see Tab.~\ref{Tab3}).}
\label{SFig1}
\end{figure*}

Table~\ref{Tab3} summarizes our model for the \EP\ coupling matrix elements in graphene
$g^\nu_{m',\,m}({\bf p},{\bf q})$ [Eq.~(\ref{eq:g})]
and we show in Fig.~\ref{Fig1} that this model, with the use of linearized
Dirac cones, can reproduce extremely well first-principles resistivity,
and the relative contributions arising from each phonon branch.
The model also accurately reproduces the details of the \EP\ coupling matrix elements
(see Figs.~\ref{SFig2} and~\ref{SFig1}).

As mentioned, there are two different contributions
to $\RHO$ from the LA phonon branch:
one arising from the gauge field and the other arising from the deformation potential
(see Tab.~\ref{Tab3} and Refs.~\cite{PhysRevB.76.045430,PhysRevB.85.165440}).
Among the two, only the deformation potential contribution depends sensitively on the
screening, or doping (see, e.\,g.\,, Ref.~\cite{PhysRevB.82.195403}).
On the other hand, the \EP\ coupling matrix elements for
the undoped case and those of heavily doped ones are almost the same
(Fig.~\ref{SFig1}),
even if screening is different;
hence, the deformation-potential contribution to $\RHO$ is
much smaller than the gauge-field contribution.
Therefore, ascribing $\RHO$ obtained from experiments to the deformation potential
alone can lead to a significant overestimation of the deformation potential extracted.
In our model, we consider only the gauge-field contribution to $\RHO$
(setting $D=0$ in Tab.~\ref{Tab3}).

Next, we study the effects of doping on the \EP\ coupling matrix elements.
Our results on doped and undoped systems
are found to be very close to each other for both the case of
acoustic phonons (Fig.~\ref{SFig1}) and that of optical and zone-boundary
(Fig.~\ref{SFig2}) phonons. Consequently, the $\RHO(T)$ that is
obtained using the coupling matrix elements of pristine graphene, but
shifting $E_{\rm F}$ in Eq.~(\ref{eq:a2tr}) appropriately, will only be a few percent
different from the $\RHO(T)$ of {\it n}-doped graphene
obtained explicitly taking into account the doping dependence of the
\EP\ coupling matrix elements.

We now incorporate \EE\ interaction effects beyond LDA
into the \EP\ coupling matrix elements.
For intra-valley scattering phonons with wavevectors near $\Gamma$,
it is known for optical phonons that
$|g^{\Gamma}_{\EE}/g^{\Gamma}_{\rm LDA}|\approx v_{\EE}/v_{\rm LDA}$~\cite{PhysRevB.77.041409,PhysRevB.78.081406},
where $g$'s and $v$'s are the corresponding \EP\ coupling matrix elements
and the Fermi velocities, respectively.
Likewise, we assume that the same relation applies to acoustic phonon branches
with wavevectors near $\Gamma$.
Since $\RHO\propto g^2/v^2$ [Eqs.~(\ref{eq:a2tr}) and~(\ref{eq:RHO})
and Tab.~\ref{Tab3}],
the contribution to $\RHO$
of the phonons with wavevectors near $\Gamma$ does not change
even when \EE\ interaction effects beyond the LDA are introduced.
On the other hand, the energy of the $A_1'$ phonons with wavevector near K
changes from $\hbar\omega^{A_1'}_{\rm LDA}=0.160$~eV to
$\hbar\omega^{A_1'}_{\EE}=0.150$~eV due to \EE\
interactions~\cite{PhysRevB.78.081406,PhysRevB.80.085423},
when treated within the {\it GW} approximation~\cite{PhysRev.139.A796}.
In addition, the \EP\ coupling matrix elements increase
when \EE\ interaction effects beyond LDA are taken into account
(see Ref.~\cite{rubio:nl2010}).
Thus, we have taken the numerical values for
$|D^{A_1'}_{\EE}/D^{A_1'}_{\rm LDA}|^2=|g^{A_1'}_{\EE}/g^{A_1'}_{\rm LDA}|^2 \cdot \omega^{A_1'}_{\EE}/\omega^{A_1'}_{\rm LDA}$
(which decreases with doping) from Ref.~\cite{rubio:nl2010}.
Considering all these effects,
together with the velocity renormalization of $v_{\EE}=1.0\times10^{6}$~m/s
and $v_{\rm LDA}=0.866\times10^{6}$~m/s,
provides a resistivity
within the {\it GW} approximation.
We show in Fig.~\ref{FigGW} that these \EE\ interaction effects beyond the LDA come into play
at high temperatures (because only the resistivity arising from high-energy $A_1'$
phonons is affected) and also decrease upon doping.

Figure~\ref{Fig3} summarizes all our final results for the resistivity
as a function of doping and temperature, and compares them with experiments.
Our results reproduce well both
the low and high temperature regimes observed~\cite{PhysRevLett.105.256805},
with theoretical data at most 30--40\,\% lower than measured values.
Importantly, again, we predict a steep increase of  
the slope $d\RHO/dT$, as a result of the strong contribution of
the optical and zone-boundary phonon modes, at temperatures higher than
those accessed in current experiments~\cite{PhysRevLett.105.256805},
suggesting the importance of further, higher temperature tests.

As mentioned earlier,
previous theoretical studies~\cite{PhysRevB.81.121412,PhysRevB.85.165440} based
on first-principles results underestimated $\RHO$ from
experiments~\cite{PhysRevLett.105.256805,chen:nnano} by
$\sim4$~times~\cite{PhysRevB.81.121412} and
$\sim13$~times~\cite{PhysRevB.85.165440}, respectively.
We attribute these discrepancies
partly to the difference in the calculated \EP\ coupling matrix elements,
and partly to the inclusion of \EE\ interaction effects beyond LDA
(e.g., in Ref.~\cite{PhysRevB.85.165440},
although the velocity enhancement due to \EE\ interactions
in the denominator of Eq.~(\ref{eq:RHO}) was considered,
the enhancement of the \EP\ coupling matrix elements
and the renormalization of the A$_1'$ phonon
frequencies~\cite{PhysRevB.78.081406,PhysRevB.80.085423}
were not considered, leading to an underestimation of $\RHO$).

Our main findings can thus be summarized as follows:
(i) The acoustic-phonon contribution to $\RHO$
of the gauge field is much more important than that of the deformation potential.
(ii) The resistivity $\RHO$ arising from the TA phonon modes is
2.5~times larger than that arising from the LA phonon modes.
(iii) The high-energy optical and zone-boundary phonons in graphene
($\omega^\nu_{\bf q}\ge0.150$~eV) are responsible for 50\,\% of $\RHO$
even at room temperature and become dominant at higher temperatures.

In conclusion, we have shown that state-of-the-art
first-principles calculations employing ultra-dense Brillouin zone sampling
accurately reproduce the
charge-density and temperature dependence of the intrinsic electrical resistivity of graphene
and provide a detailed microscopic understanding
of the relative role of different phonon modes.
Moreover, we have shown that it is possible to build an analytical model for the \EP\ interactions
that retains the accuracy of first-principles calculations:  this model represents a useful
reference for fundamental studies of carrier dynamics in low-dimensional graphitic systems as well
as a tool for graphene-based electronic devices simulations.

{\bf Key words:} graphene, electron-phonon interaction,
intrinsic electrical resistivity,
deformation potential, gauge field, {\it GW} approximation.

{\bf Acknowledgment:} CHP acknowledges support from Korean NRF funded by MSIP
(Grant No.\,NRF-2013R1A1A1076141), NB from EU FP7/CIG Grant No.\,294158,
MC, FM and TS from ANR-11-IDEX-0004-02, ANR-11-BS04-0019, ANR-13-IS10-0003-01,
and the Graphene Flagship, GS and BK from US NSF under Grant No.\,1048796,
and NM from Swiss NSF 200021$\_$143636.
Computer facilities were provided by PLSI of KISTI, CSCS, CINES, and IDRIS.

\end{document}